# MATHEMATICAL MODELS OF SPECIFIC ELEMENTS OF WIND ENERGY CONVERSION SYSTEMS BASED ON INDUCTION GENERATOR

**Octavian PROSTEAN, Iosif SZEIDERT, Nicolae BUDISAN,
Gabriela PROSTEAN, Cristian VASAR**

Department of Automation and Industrial Informatics
Faculty of Automation and Computer Science
Bd. V. Parvan, No.2, 1900 Timisoara, ROMANIA
Tel: +40 256 403237  Fax: +40 256 403214  E-mail: siosif@aut.utt.ro

*Abstract:* In the complex case of wind energy conversion systems (WECS) analysis and synthesis, the mathematical models of each component element are required. In the paper are presented some contributions to preset values establishing strategies and models of specific elements: rotation speed and voltage preset values establishing strategies, a simplified fixed blade wind turbine model, the induction generator squirrel rotor induction generator Park mathematical model, specific to current frequency converter supplier use, current frequency converter inverters' model. The presented models are used for WECS analysis in its specific transient regimes as well for its controllers' design.

## 1. Introduction

There are known, presently, a large number of structure variants for induction machine field-oriented control. The considerations presented in the paper are standing for the most of those structures. In the following, for presentation, will be considered only the case of the induction machine air-gap magnetic field-oriented control system structure. (As illustrated in figure 1)

The structure is identical as in the case of electrical drives, in which appear: transducers /estimators, controllers and the F. Blaschke field-oriented method specific calculator blocks (phase number transformers TS, $TS^{-1}$, field orientation block FOB and its elements, phase analyzer PHA, axes transformers AT, $AT^{-1}$), frequency converters, excepting the PRESET TURBINE - GENERATOR ROTOR SPEED CALCULATOR, PRESET VOLTAGE CALCULATOR and WIND TURBINE-INDUCTION GENERATOR GROUP. [2]

There are studied the specific windmill's control systems elements.

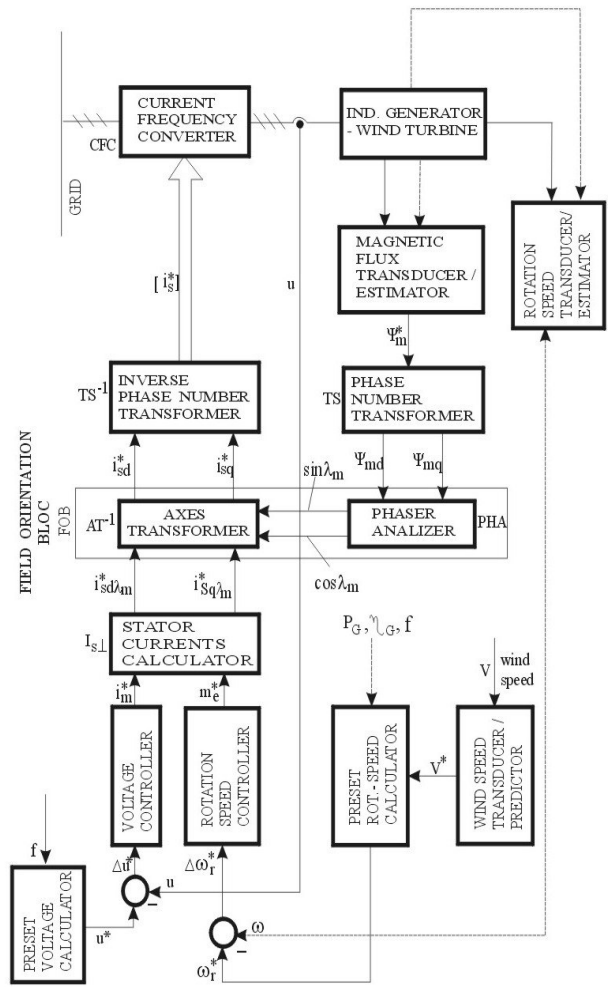

Figure 1. Induction machine air-gap magnetic field-oriented control system structure.

## 2. Mathematical Models of Specific Elements

### 2.1. Preset Turbine-Generator Rotor Speed Calculator

It is known that for each wind speed an optimum rotation turbine speed is desired. Based on original turbine mathematical model (see paragraph 2.4), a sensorless rotation speed determination strategy was elaborated; correspondingly a rotation speed calculation method was proposed [1].

### 2.2. Preset Generator Voltage Calculator

In case of the variable speed induction machine, optimum operation may be assured [3] if

$$U = U_N \cdot \frac{f}{f_N} \sqrt{\frac{M}{M_N}} \quad (1)$$

(in the case of neglecting the magnetic saturation).
In the case of coupled generator machines with ventilators and wind turbines, having $M \equiv n^2 \equiv f^2$ the relation (1) becomes

$$U = U_N (\frac{f}{f_N})^2 \quad (2)$$

This fact must be considered for variable speed induction generator windmills.
As a simplification, in the literature, there is admitted that,

$$U = U_N \cdot \frac{f}{f_N} \quad (3)$$

Correspondingly to (2), the induction generator preset voltage calculator must accomplish the task presented in figure 2.

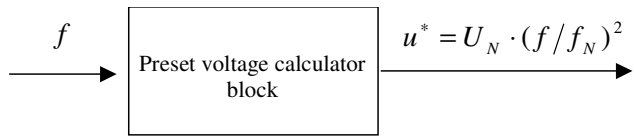

Figure 2. The induction generator preset voltage calculator.

### 2.3. Frequency Converter Models

There are different frequency converters: thyristor, transistor, cyclo-converter, intermediate direct current circuit, matrix (latest achievements). The modern implementations are based on IGBT transistor PWM converters with direct current intermediate circuit. The majority of models may be taken into account for the three-phase inverters of considered converters. A very simple model may be considered as a device with three static switches, one for each phase. The IGBT commutation times are neglected. The simplified scheme of inverter is presented in figure 3.

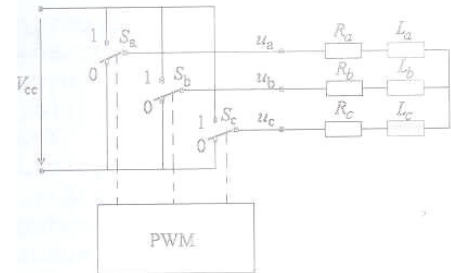

Figure 3. Inverter simplified model.

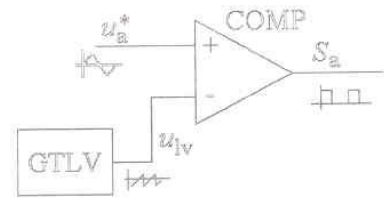

Figure 4. PWM electric scheme.

Regarding the figure 3, the inverter's mathematical model is:

$$u_a = \frac{1}{3} V_{cc} (2S_a - S_b - S_c)$$
$$u_b = \frac{1}{3} V_{cc} (-S_a + 2S_b - S_c) \quad (4)$$
$$u_c = \frac{1}{3} V_{cc} (-S_a - S_b - S_c)$$

Where: $S_i (i = a, b, c)$ are control signals delivered by the PWM circuit, that have the values 0 or 1, depending on the fact that switch element is "on" or "off". There was elaborated a Spice program for the frequency converter (figure 3) and for the PWM generator (figure 4). The simulation results concludes that for small load (current) the distortion of current wave is greater as in the case of a higher value load. The Canadian HEXATRONIC Inc., Toronto research partner in the domain of frequency converters, at his converter test stand:
· Input DC Voltage (buck): 460 VDC
· Nominal current: 10 ADC
· Output voltage: 208 VAC,
at low load current of 3 Aef AC, obtained [7], for the inverter, the current and voltage waveforms shown in figure 5 with a satisfactory current harmonic distortion factor THD of 5.23% (considering the IEEE 519 harmonic distortion standard requirements for considered load conditions: THD<5). In figure 6, there is presented the HEXATRONIC Inc. power electronic converter implementation.

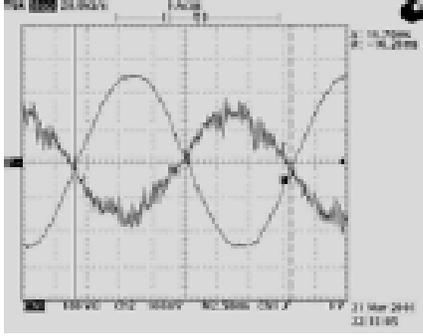

Figure 5. Line voltage and current waveforms obtained by HEXATRONIC converter test stand (in above presented load conditions).

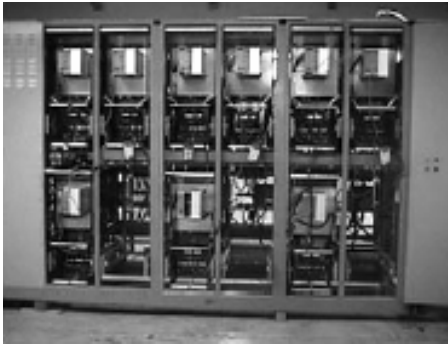

Figure 6. HEXATRONIC implementation: 3000HP @ 4160V power electronic converter.

### 2.4. Fixed Blade Wind Turbine Models

The mathematical models presented in this paper consider the case of the fixed blade wind turbines and expresses the mathematical power and moment driving the electrical generator. The established models contain the evaluation of all kind of factors: turbine type, dimensions, rotation speed, and wind parameters. The usual blade power and moment models are considered:

$$P_{uT} = C_{P_{uT}} \cdot \rho \cdot (v^3/2) \cdot S_T \quad (5)$$

$$M_{uT} = P_{uT}/\omega_T = C_{M_{uT}} \cdot \rho \cdot (v^2/2) \cdot S_T \cdot R_T \quad (6)$$

Where: $C_{P_{uT}}, C_{M_{uT}}$ - blade useful power and torque coefficients, $\rho$ - air density, v – wind speed, $S_T$ – turbine swept aria, $R_T$ – turbine radius.

For power and torque coefficients there are proposed the following expressions: [2]

$$C_{P_{uT}} = C_{M_0} \cdot \lambda + a\lambda^\alpha - b\lambda^\beta \quad (7)$$

$$C_{M_{uT}} = C_{M_0} + a\lambda^{\alpha-1} - b\lambda^{\beta-1} \quad (8)$$

Where: $C_{M_0}$ – start torque coefficient, $\lambda = u/v$ - type speed ratio, u – blade peripheral, $\alpha, \beta, a, b$ – turbine blade constant values.

The constants $C_{M_0}, \alpha, \beta, a, b$ depend on turbine type and construction. They were established and expressed as a result of analysis of a large number of world wide constructed turbines. On this basis, the following expressions were determined:

$$(C_{P_{uT}})_{max} = 0.3 \ \lambda_0^{0.35} - 0.0014\lambda_0^2 \quad (9)$$

$$C_{M_0} = 0.2/\lambda_0^2 \quad (10)$$

$$\alpha = 2...2.5 \quad (11)$$

$$\beta = 2.3...2.8 \quad (12)$$

$$b = (C_2/A_2 - C_1/A_1)/(B_1/A_1 - B_2/A_2) \quad (13)$$

$$a = B_2/A_2 \cdot b + C_2/A_2 \quad (14)$$

$$A_1 = \lambda_0^\alpha \quad (15)$$

$$A_2 = \alpha \cdot \lambda_0^{(\alpha-1)} \quad (16)$$

$$B_1 = \lambda_0^\beta \quad (17)$$

$$B_2 = \beta \cdot \lambda_0^{(\beta-1)} \quad (18)$$

$$C_1 = (C_{P_{uT}})_{max} - \lambda_0 \cdot C_{M_0} \quad (19)$$

$$C_2 = -C_{M_0} \quad (20)$$

There can be noticed, that in order to particularize the model the optimal type speed ratio $\lambda_0$ must be considered.

The simplification of proposed above models may be continued as a result of some considerations on the values that they contain. Therefore, at turbines with high optimal type speed ratio $\lambda_0$, can be neglected the influence of $C_{M_0}$ at the turbine start. Considering, in this case, $C_{M_0} = 0$, there results:

$$C_{P_{uT}} = a\lambda^\alpha - b\lambda^\beta \quad (21)$$

$$P_{uT} = \rho \cdot (v^3/2) \cdot S_T \cdot (a\lambda^\alpha - b\lambda^\beta) \quad (22)$$

or $\quad P_{uT} = \rho \cdot (v^3/2) \cdot S_T \cdot (a(\frac{u}{v})^\alpha - b(\frac{u}{v})^\beta) \quad (23)$

or $\quad P_{uT} = \rho \cdot (\frac{v^3}{2}) \cdot S_T (a(\frac{\omega_r R_T}{v})^\alpha - b(\frac{\omega_r R_T}{v})^\beta) \quad (24)$

and

$$M_{uT} = \rho \cdot (\frac{v^2}{2}) \cdot S_T \cdot R_T (a \cdot (\frac{\omega_r R_T}{v})^{\alpha-1} - b(\frac{\omega_r R_T}{v})^{\beta-1}) \quad (25)$$

## 2.5. "Wind Turbine - Induction Generator" Mathematical Model

The usage of the turbine (25) and the generator (Park-Kron) models [2], for the turbine-induction generator direct coupled group results the following relation:

$$J_{T,G}\frac{d\omega_r}{dt}=|M_{uT}|-|M_{e,G}| \qquad (26)$$

where: $M_{uT}$ – useful wind turbine moment (25), $M_{e,G}$ - generator moment, resulting from its exposed below mathematical model.
The induction generator transient equations (27-31):

$$\frac{dI_d}{dt}=-\frac{L_2 R_1}{L_1 L_2 - M^2}\cdot I_d + \frac{\omega L_1 L_2}{L_1 L_2 - M^2}\cdot I_q + \frac{M\omega L_2}{L_1(L_1 L_2 - M^2)}\cdot I_{qr} +$$

$$+\frac{L_2}{L_1 L_2 - M^2}U_d - \frac{M^2(\omega-\omega_M)}{L_1 L_2 - M^2}\cdot s\cdot I_q$$

$$+\frac{R_2 M}{L_1 L_2 - M^2}\cdot I_{dr} - \frac{ML_2(\omega-\omega_M)}{L_1 L_2 - M^2}\cdot s\cdot I_{qr}$$

$$\frac{dI_q}{dt}=-\frac{\omega L_1 L_2}{L_1 L_2 - M^2}\cdot I_d - \frac{R_1 L_2}{L_1 L_2 - M^2}\cdot I_q - \frac{ML_2\omega}{L_1 L_2 - M^2}\cdot I_{dr} +$$

$$+\frac{M^2(\omega-\omega_M)}{L_1 L_2 - M^2}\cdot s\cdot I_d + \frac{ML_2(\omega-\omega_M)}{L_1 L_2 - M^2}\cdot s\cdot I_{dr} + \frac{R_2 M}{L_1 L_2 - M^2}\cdot I_{qr}$$

$$\frac{dI_{dr}}{dt}=\frac{MR_1}{L_1 L_2 - M^2}\cdot I_d - \frac{L_1 M\omega}{L_1 L_2 - M^2}\cdot I_q - \frac{M^2\omega}{L_1 L_2 - M^2}\cdot I_{qr} -$$

$$-\frac{M}{L_1 L_2 - M^2}U_d + \frac{ML_1(\omega-\omega_M)}{L_1 L_2 - M^2}\cdot s\cdot I_q -$$

$$-\frac{R_2 L_1}{L_1 L_2 - M^2}\cdot I_{dr} + \frac{L_1 L_2(\omega-\omega_M)}{L_1 L_2 - M^2}\cdot s\cdot I_{qr}$$

$$\frac{dI_{qr}}{dt}=\frac{L_1 M\omega}{L_1 L_2 - M^2}\cdot I_d + \frac{MR_1}{L_1 L_2 - M^2}\cdot I_q + \frac{M^2\omega}{L_1 L_2 - M^2}\cdot I_{dr} -$$

$$-\frac{ML_1(\omega-\omega_M)}{L_1 L_2 - M^2}\cdot s\cdot I_d - \frac{L_1 L_2(\omega-\omega_M)}{L_1 L_2 - M^2}\cdot s\cdot I_{dr} - \frac{L_1 R_2}{L_1 L_2 - M^2}\cdot I_{qr}$$

$$\frac{ds}{dt}=\frac{1}{2\pi}\cdot M\cdot I_d\cdot I_{qr} - \frac{1}{2\pi}\cdot M\cdot I_q\cdot I_{dr} + \frac{1}{2\pi}\cdot M_{rez} \quad (27\text{-}31)$$

Where: $L_1$ - stator winding inductance, $L_2$ - rotor winding inductance, $R_1$ - rotor winding resistance, $R_2$ - rotor winding resistance, $\omega$ - electric angular rotation speed, M - mutual stator-rotor inductance, $U_d$ - line voltage, n - rotation speed, $I_d(t)$ – stator d-axis current, $I_q(t)$ – stator q-axis current, $I_{dr}(t)$ – rotor d-axis current, $I_{qr}(t)$ – rotor q-axis current, s(t) = s(t) – slip

## 2.6. Induction Generator Simulation Issues.

The equation system (27-31) was implemented in Matlab-Simulink environment, resulting the schematic diagram presented in figure 7.

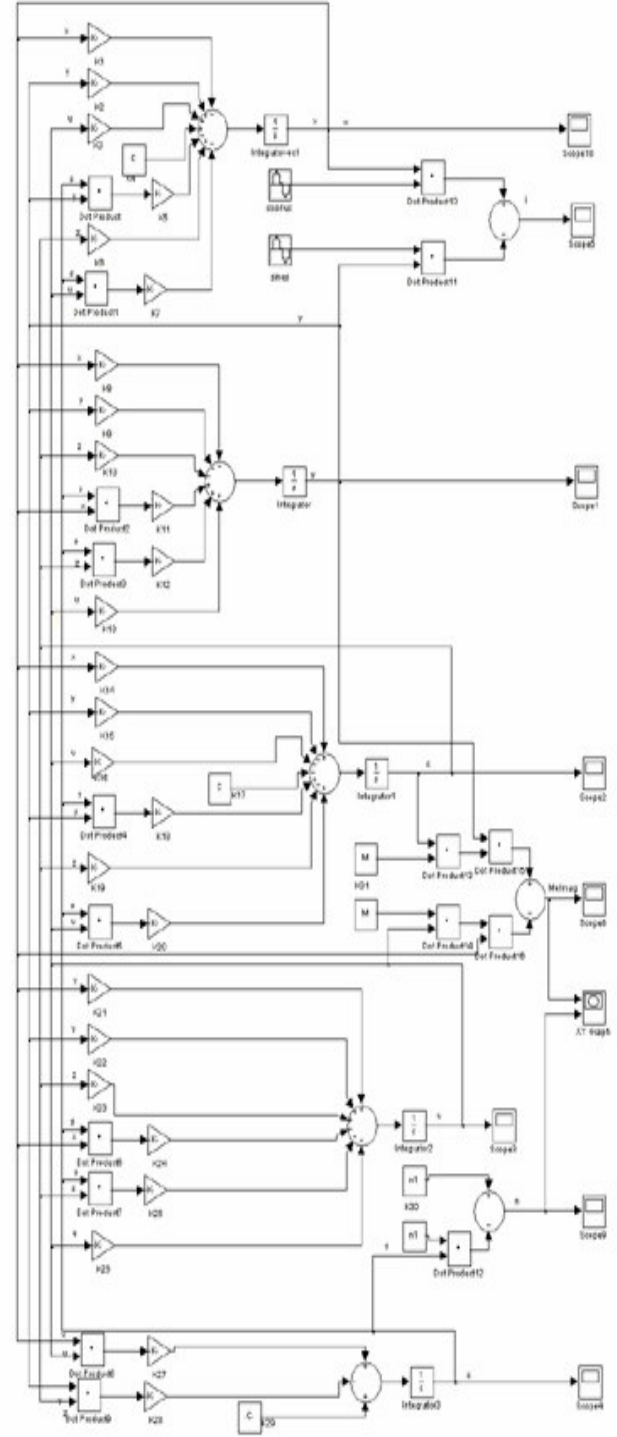

Figure 7. Induction Generator - Simulink schematic

diagram.

In order to exemplify the usage of the above presented model, in figures 8 and 9 are presented the obtained results in the case of three-phased short-circuit simulation. [5],[6] The current evolution is represented in [A], t[sec] axes and the torque evolution in [Nm], t[sec] axes.

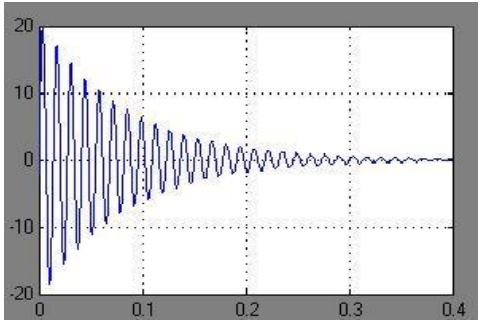
Figure 8. Induction generator current evolution.

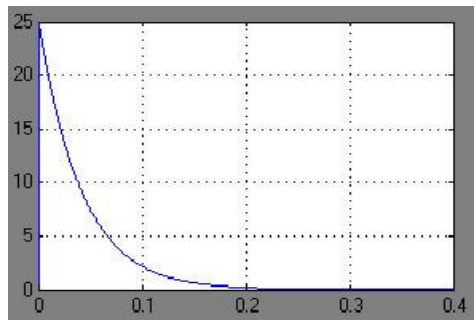
Figure 9. Induction generator torque evolution

## 3. Conclusions

In the paper there are presented considerations on mathematical models of specific elements of induction generator wind energy conversion systems. [2],[4],[8]

The obtained models can be used independently, in order to solve transient processes in specific operation regimes: generator short-circuit, change of turbine moment due to wind gusts, change of grid voltage, generator grid connection, change of wind mill load, and other particular regimes of interest.

Together with mathematical models of all elements, controllers inclusively, may be obtained the complete simulation scheme of the entire wind energy conversion control system and, correspondingly, the above-mentioned problems may be solved in these circumstances.

The research results, of a National University Research Council project [3], concerns also the use of mathematical model of conversion system, for controllers' simulation design.